\documentclass[preprint,showpacs]{revtex4}
\usepackage{epsfig}
\usepackage{graphicx}
\usepackage{bm}
\usepackage{amsmath} 
\newcommand{\half}{\mbox{${\textstyle \frac{1}{2}}$}}           % 1/2
\newcommand{\rd}{\textrm{d}}
\begin{document}
%{\flushright{\today}}
\title{Polarization observables in the $e^+ e^- \rightarrow   \bar{\Lambda} \Lambda$ reaction}
\date{\today}
\author{G\"oran F\"aldt}\email{goran.faldt@physics.uu.se}  
\affiliation{ Department of physics and astronomy, \
Uppsala University,
 Box 516, S-751 20 Uppsala,Sweden }

\begin{abstract}
Cross-section, vector-polarization, and tensor-polarization distributions are calculated 
for the reactions $e^+ e^- \rightarrow   \bar{p}p$ and 
$e^+ e^- \rightarrow   \bar{\Lambda} \Lambda$. Each reaction requires six characteristic functions
that are bilinear in the, possibly complex, electromagnetic form factors, denoted 
$G_E(P^2)$  and $G_M(P^2)$, of $p$ and $\Lambda$. 
For the hyperon reaction also the
joint-decay distributions of $\Lambda$ and $\bar{\Lambda}$ are calculated. 
Their knowledge allow a complete determination of the hyperon electromagnetic form factors,
without measuring hyperon spins. We explain
how this is done in practice. For some tensor-polarization components our
results are in conflict with previously repeatedly published distributions.

\end{abstract}
\maketitle

%
%%%%%%%%%%%%%%%%%%%%%%%%%%%
%
\section{Introduction}\label{ett}

The cross-section distributions of the reaction $e^- e^+ \rightarrow  p \bar{p} $ or 
$\Lambda \bar{\Lambda}$ are governed by two form factors  $G_E(P^2)$  and $G_M(P^2)$,
with time-like argument $P^2$. The unpolarized distribution is proportional to a 
certain linear combination of  $|G_E|^2$  and $|G_M|^2$. The combination 
 $\textrm{Im} (G_MG_E^\star )$ is proportional to the only non-vanishing component 
of the vector polarization $P_y$. An independent linear combination of $|G_E|^2$  and $|G_M|^2$
can be obtained from any of the diagonal components of the tensor polarization $A_{ii}$.
The only non-vanishing non-diagonal components of the tensor polarization, $A_{xz}=A_{zx}$
are proportional to  $\textrm{Re} (G_MG_E^\star )$. For a complete determination of the
form factors these four quantities must be measured.

The reactions $e^- e^+ \rightarrow  p \bar{p} $ and $\Lambda \bar{\Lambda}$ are presently under 
investigation at BESIII, the Beijing Spectrometer III. 
Details of the analysis and mesurement of such reactions are given in the Varenna lectures 
of Johansson \cite{TJV}.
Expressions for the vector and tensor polarizations    were
first calculated by   Dubni\v{c}kova {\itshape et al}.\ \cite{DDR}. Unfortunately, this reference has 
many misprints. Corrected expressions are given by Gakh and Tomasi-Gustafsson \cite{GTG}.
The expressions of these authors are confirmed by Buttimore and Jennings \cite{BeJ}. 
The reason for writing this paper is that we believe there are still errors in the 
last two publications. In contrast to their conclusion we claim that also the 
non-diagonal components $A_{yx}=A_{xy}$ vanish, leaving $A_{xz}=A_{zx}$ as the only 
non-vanishing non-diagonal components.

In addition to calculating the vector and tensor polarizations
we calculate the hyperon-decay distributions in the reaction 
$e^- e^+ \rightarrow  \Lambda(\rightarrow p\pi^-) \bar{\Lambda}(\rightarrow \bar{p}\pi^+)$.
For this pupose we use the folding method of Czy\.z {\itshape et al}.\ \cite{Czyz},
but in the covariant version developed in ref.\ \cite{GF}, with proper counting
of  intermediate states. This application requires knowlege of the cross-section distributions for
$e^- e^+ \rightarrow  \Lambda \bar{\Lambda}$ with polarized hyperons.

%\newpage
%
%%%%%%%%%%%%%%%%%%%%%%%%%%%
%
\section{Lambda form factors}\label{två}

The reaction under consideration is described by the diagram of fig.1. Momentum definitions
are also indicated there. The couplings of the
initial state leptons are  determined by the electron charge. Form factors or 
anomalous magnetic moments  are not considered.
\begin{figure}[ht]
\begin{center}
%\resizebox{0.38\textwidth}{!}
\scalebox{0.50}{ \includegraphics{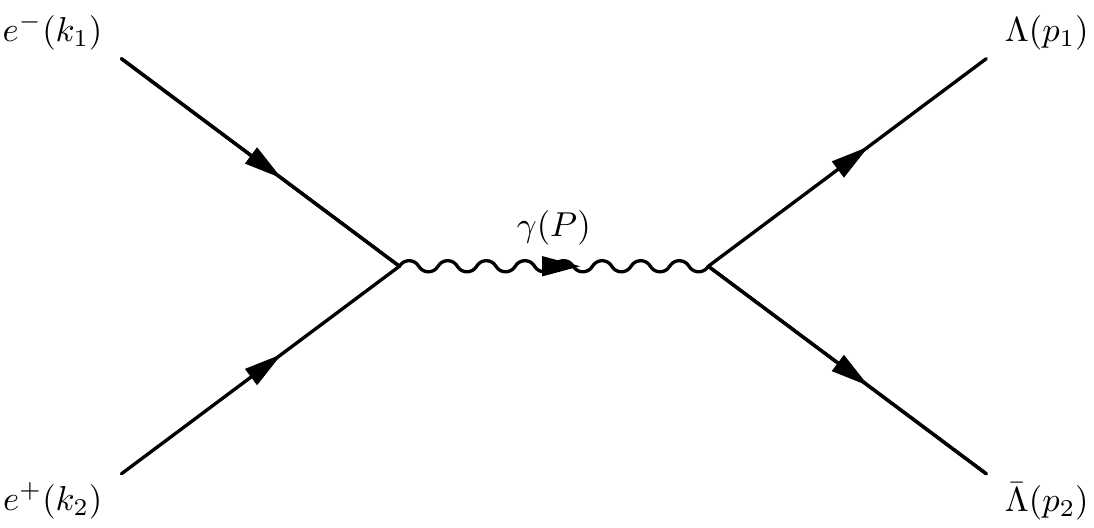} }    
\caption{Graph describing the reaction 
$e^+ e^- \rightarrow  \bar{\Lambda} \Lambda$.}
\label{F1-fig}
\end{center}
\end{figure}

In the current matrix elements of the lambda, however, both form
factors  are taken into account.
We follow common practice \cite{GF} and  write 
the hadron current matrix element as
\begin{eqnarray}
		j_\mu(p_1,p_2)&=&-ie\bar{u}(p_1)O_\mu(p_1,p_2)v(p_2), \\
		O_\mu(p_1,p_2)&=&G_1(P^2)\gamma_\mu-
		  \frac{1}{2M}G_2(P^2)Q_\mu , \label{Lambdavertex}
\end{eqnarray}
with $P=p_1+p_2$ and $Q=p_1-p_2$. The lambda mass is denoted $M$.

The form factors $G_1$ and $G_2$ are related to the more commonly used form factors $F_1$ and $F_2$,
and the electric $G_E$ and magnetic 
$G_M$ form factors \cite{Czyz,Pil,Dalk}, through
\begin{eqnarray}
	G_1&=& F_1+F_2 = G_M	  \\
	G_2&=&F_2=\frac{1}{1+\tau}(G_M-G_E) =\frac{4M^2}{Q^2}(G_M-G_E), \qquad
\end{eqnarray}
and $\tau=-P^2/4M^2$. The arguments of the form factors are all equal to $P^2$. In particular, when 
$P^2=4M^2$ then $G_M=G_E=F_1+F_2$. Another useful relation is
\begin{equation}
	Q^2=-4 p^2,
\end{equation}
where $p$ is the final-state c.m.\ momentum.
%\newpage
%
%%%%%%%%%%%%%%%%%%%%%%%%%%%
%
\section{Cross section}\label{tre}

We follow Pilkuhn \cite{Pil} and write the cross-section distribution of the reaction
$e^+ e^- \rightarrow \bar{\Lambda} \Lambda $ as
\begin{equation}
	\rd \sigma= \frac{1}{2\sqrt{\lambda(s,m_e^2,m_e^2)}} \, \overline{|{\cal{M}}|^2}\,
	   \textrm{dLips}(k_1+k_2;p_1,p_2)	  ,\label{dSig1}
\end{equation}	   
where the average over the squared matrix element indicates 
 average over initial electron and positron spins.  The definitions of the
particle momenta follows fig.\ 1.

We  remove some trivial factors from the squared matrix element,
namely the powers of the electron charge and the square of the intermediate-photon
denominator, and write
\begin{equation}
	\overline{|{\cal{M}}|^2} = \left(	\frac{e^2}{P^2} \right)^2 \overline{|{\cal{M}}_{red}|^2}\ .
	 \label{dSig2}
\end{equation}	
The reduced matrix element is decomposed as
\begin{equation}
	\overline{|{\cal{M}}_{red}|^2} =  L_{\nu\mu} K^{\nu\mu}\ , \label{dSig3}
\end{equation}	
where $L_{\nu\mu}$ and $K_{\nu\mu}$ are the lepton and hadron electromagnetic tensors.
%

%
%%%%%%%%%%%%%%%%%%%%%%%%%%%
%\newpage
%
\section{Lepton tensor}\label{fyra}
The lepton tensor is by definition equal to  
\begin{equation}
	L_{\nu \mu}(k_1,k_2)=\frac{1}{4}\mbox{Sp}[\gamma_\nu (\not\! k_2 - m_e )\gamma_\mu (\not\! k_1 +m_e ) ]   \ ,
	 \label{Lepton_vector}
\end{equation}
and takes care of the average over initial-state-lepton spins.
 We shall neglect the electron mass  $m_e$ compared with other masses 
and energies. In this approximation
\begin{eqnarray}
	L_{\nu\mu}& = & L_{\mu\nu} \nonumber   \\
	 &=& k_{1\nu}k_{2\mu}+ k_{2\nu}k_{1\mu} -\half s 
	    g_{\nu\mu},\label{Lepton_tensor}
\end{eqnarray}
with
\begin{equation}
	s  =  (k_1+k_2)^2 = P^2.           
\end{equation}

The lepton tensor enters  the cross-section distribution contracted
with the hadron tensor. The hadron tensor is gauge invariant, which means that when contracted
with four vectors $P^\mu$ or $P^\nu$ zero result is obtained. Hence, dependencies 
$P_\mu$ or $P_\nu$  in the lepton
tensor may be ignored, and as an example, we may replace eq.\ (\ref{Lepton_tensor}) by
\begin{equation}
	L_{\nu\mu} = -2 k_{1\nu}k_{1\mu} -\half s 
	    g_{\nu\mu}\ .
\end{equation}

%
%
%%%%%%%%%%%%%%%%%%%%%%%%%%%
%\newpage
%
\section{Hadron tensor}\label{fem}

The hadron tensor is calculated in Appendix B of ref\ \cite{GF}. 
We simply copy the formulae given there. Starting from the hadron vertex $O_\mu(p_1, p_2)$
of eq.\ (\ref{Lambdavertex}) we get 
\begin{align}
	K_{\nu\mu}(s_1,s_2)= \mbox{Sp}[& \bar{O}_\nu (\not\! p_1+M)\half(1+\gamma_5 \not\! s _1) \nonumber \\
	  & \quad \times O_\mu  (\not\! p_2-M)\half(1+\gamma_5 \not\! s _2) ], \label{LLbar-tensor}
\end{align}
with $\bar{O}=\gamma_0 O^\dagger \gamma_0$. The four vectors $s_1$ and $s_2$
denote the spin vectors of hyperon and anti-hyperon.

The hadronic tensor is gauge invariant, i.e.\ vanishes when contracted by $P^\nu$ or $P^\mu$, 
and is decomposed as
\begin{eqnarray}
	K_{\nu\mu}(s_1,s_2)&=&K_{\nu\mu}^{00}(0,0) + K_{\nu\mu}^{05}(s_1,0) + K_{\nu\mu}^{50}(0,s_2) \nonumber \\
	 &&+ K_{\nu\mu}^{55}(s_1,s_2).
	\label{LLtensdiv}
\end{eqnarray}
The functional arguments indicate the spin vectors involved. 

The explicit expression for the first part of the hadron tensor  is 
	\begin{eqnarray}
	K_{\nu\mu}^{00}&=&\frac{1}{2} \bigg( P_\nu P_\mu -P^2g_{\nu\mu}- Q_\nu Q_\mu  \bigg)|G_1|^2 \nonumber \\
	   &&+\frac{1}{2} Q_\nu Q_\mu\bigg(2\textrm{Re} (G_1 G_2^{\star})
	  -\frac{Q^2}{4M^2}|G_2|^2\bigg).  \qquad \label{hOO}  
\end{eqnarray}

 Since the lepton tensor is symmetric
in its indices we need only retain the symmetric parts of the hadron tensor. It follows that
\begin{eqnarray}
	K_{\nu\mu}^{05}(s_1,0) &=& \frac{-1}{2M}\textrm{Im}(G_1G_2^\star) \bigg[Q_\nu\epsilon(p_1,p_2,s_1)_\mu 
	          \nonumber \\ 
	&&  \qquad \qquad
	    +Q_\mu\epsilon(p_1,p_2,s_1)_\nu \bigg] ,  \\
	K_{\nu\mu}^{50}(0,s_2) &=& \frac{-1}{2M}\textrm{Im}(G_1G_2^\star) 
	   \bigg[Q_\nu\epsilon(p_1,p_2,s_2)_\mu \nonumber \\
	   && \qquad \qquad +Q_\mu\epsilon(p_1,p_2,s_2)_\nu \bigg]    ,
\end{eqnarray}
with the epsilon-function combinations defined as 
\begin{eqnarray}
	\epsilon(p_2p_1l_1)_\nu&=&\epsilon_{\alpha\beta\gamma\nu}p_2^\alpha p_1^\beta l_1^\gamma,\label{epsI} \\
	\epsilon(p_2p_1)_{\nu\mu}&=&\epsilon_{\alpha\beta\nu\mu}p_2^\alpha p_1^\beta ,
\end{eqnarray}
and $\epsilon_{0123}=1$.

The  contribution depending on both spin vectors is  
\begin{eqnarray}
K_{\nu\mu}^{55}(s_1,s_2)&=&|G_1|^2B^{1}_{\nu\mu} +|G_2|^2B^{2}_{\nu\mu} \nonumber \\ 
   && +\textrm{Re}(G_1G_2^\star) B^{3}_{\nu\mu} ,
\end{eqnarray}
with
\begin{eqnarray}
	B^{1}_{\nu\mu}&=&-s_1\cdot s_2 \bigg[ p_{1\nu} p_{2\mu}  +p_{1\mu}  p_{2\nu} -\half g_{\nu\mu}P^2 \bigg] 
	 \nonumber \\ &&
	    -\half P^2 ( s_{1\nu}   s_{2\mu} + s_{1\mu}   s_{2\nu} )  
				 -g_{\nu\mu}p_1\cdot s_2    p_2\cdot s_1 \nonumber \\
	     &&+p_1\cdot s_2 ( s_{1\nu}p_{2\mu}+s_{1\mu}p_{2\nu}) \nonumber \\ &&
	     +p_2\cdot s_1 ( s_{2\nu}p_{1\mu}+s_{2\mu}p_{1\nu})
	            ,\\
	B^{2}_{\nu\mu}&=& \frac{1}{4M^2} Q_\nu Q_\mu \bigg[ \half Q^2 s_1\cdot s_2+p_1\cdot s_2 p_2\cdot s_1 \bigg],\\
   	B^{3}_{\nu\mu}&=& -  Q_\nu Q_\mu  s_1\cdot s_2 +
      \half \bigg[ p_1\cdot s_2( Q_\nu  s_{1\mu} +Q_\mu  s_{1\nu}) \nonumber \\
			&&- p_2\cdot s_1
          (Q_\nu  s_{2\mu} +  Q_\mu s_{2\nu}) \bigg].       
\end{eqnarray}

%
%
%%%%%%%%%%%%%%%%%%%%%%%%%%%%%%%%%%
 % \newpage
  \section{Polarization variables}

Our polarization variables ${\cal{P}}^{ab}$	are defined as follows.
For unpolarized final-hadron states the factor $\overline{|{\cal{M}}_{red}|^2}$ of 
eq.\ (\ref{dSig3}) becomes 
\begin{eqnarray}
	{\cal{P}}^{00}&=&\sum_{\pm s_1,\pm s_2} L\cdot K(s_1,s_2) \nonumber\\
	  &=& 4L\cdot K^{00}(0,0).  \label{P00}
	\end{eqnarray}
Correspondingly, for polarized hyperon and unpolarized anti-hyperon 
\begin{eqnarray}
	{\cal{P}}^{05}&=&\sum_{\pm s_2} L\cdot K(s_1,s_2) - L\cdot K(-s_1,s_2) \nonumber\\
	  &=& 4L\cdot K^{05}(s_1,0)\ ,\label{P05}
	\end{eqnarray}
whereas for unpolarized hyperon and polarized anti-hyperon
\begin{eqnarray}
	{\cal{P}}^{50}&=&\sum_{\pm s_1} L\cdot K(s_1,s_2) - L\cdot K(s_1,-s_2) \nonumber\\
	  &=& 4L\cdot K^{50}(0, s_2)\ .\label{P50}
	\end{eqnarray}
Finally, in the joint- or tensor-polarization case
\begin{eqnarray}
	{\cal{P}}^{55}&=& L\cdot K(s_1,s_2) - L\cdot K(s_1,-s_2) \nonumber\\
	               &&+ L\cdot K(-s_1,-s_2) - L\cdot K(-s_1,s_2)\nonumber\\
	  &=& 4L\cdot K^{55}(s_1, s_2)\ .\label{P55}
	\end{eqnarray}
	
	The spin four-vector $s=s(p,n)$ satisfies $s\cdot s=-1$ and $s\cdot p=0$, 
where $p$ is the hadron four-momentum. The mass of the hadron is $M$, and
its  spin vector $-s(p,n)$ for polarisation  $-\mathbf{n}$. 

In the rest system of the hadron $s(p,\mathbf{n})=(0,\mathbf{n})$ and 
$p=(M,\mathbf{0})$. In a coordinate system where the hadron has
three-momentum $\mathbf{p}$, the spin vector is
\begin{equation}
	s(\mathbf{p},\mathbf{n})=\frac{n_{\parallel}}{M}(|\mathbf{p}|, E \hat{\mathbf{p}})+(0,\mathbf{n}_\bot),
	\label{spin-vector}
\end{equation}
with $n_{\parallel}=\mathbf{n}\cdot\hat{\mathbf{p}}$ and
\begin{equation}
	\mathbf{n}_\bot=\mathbf{n} -\hat{\mathbf{p}}(\mathbf{n}\cdot \hat{\mathbf{p}}).  
\end{equation}
In the first part of expression (\ref{spin-vector}) we notice the helicity
vector $h(p)=(|\mathbf{p}|, E \hat{\mathbf{p}})/M$.

We have evaluated these polarization variables in the global c.m. system, where
\begin{eqnarray}
	\mathbf{p}_1&=& -\mathbf{p}_2 = \mathbf{p} , \\
	  \mathbf{k}_1&=& - \mathbf{k}_2 = \mathbf{k} ,
	\end{eqnarray}
and with scattering angle
\begin{equation}
	\cos(\theta )= \hat{\mathbf{p}} \cdot \hat{\mathbf{k}}.
\end{equation}
The polarization vectors of the final-state hadrons are $\mathbf{n}_1$ and 	$\mathbf{n}_2$
in their respective rest systems (cf eq.\ (\ref{spin-vector})).

It is convenient to express the polarization variables in terms of the form-factor combinations
\begin{eqnarray}
	D_c&=& 2 s |G_M|^2 , \\
	  D_s&=&  s |G_M|^2 -4M^2 |G_E|^2.
	\end{eqnarray}

The unpolarized variable becomes
\begin{equation}
	{\cal{P}}^{00}= s \left( D_c - D_s \sin^2(\theta)\right), \label{P00fin}
\end{equation}
and the vector polarization
\begin{equation}
	{\cal{P}}^{05}= {\cal{S}} \left[ \frac{1}{\sin(\theta)} ( \hat{\mathbf{p}}\times \hat{\mathbf{k}})\cdot \mathbf{n}_1 
	\right],\label{P05fin}
\end{equation}
with the  function ${\cal{S}}$ defined as
\begin{equation}
	{\cal{S}}=-2Ms \sqrt{s} \sin(2\theta) \textrm{Im} (G_MG_E^\star ). \label{DefS}
\end{equation}
To get ${\cal{P}}^{50}$ from (\ref{P05fin}) we replace $\mathbf{n}_1$ by $\mathbf{n}_2$, 
the variable ${\cal{S}}$ remaining the same,
\begin{equation}
	{\cal{P}}^{50}= {\cal{S}} \left[ \frac{1}{\sin(\theta)} ( \hat{\mathbf{p}}\times \hat{\mathbf{k}})\cdot \mathbf{n}_2 
	\right]. \label{P50fin}
\end{equation}

The joint or tensor  polarization, finally, can be written as
\begin{eqnarray}
	{\cal{P}}^{55}&=& {\cal{T}}_1 \mathbf{n}_{1}\cdot \hat{\mathbf{p}}\   \mathbf{n}_{2}\cdot \hat{\mathbf{p}}  +  
	    {\cal{T}}_2 \mathbf{n}_{1\bot} \cdot \mathbf{n}_{2\bot}  
	  + {\cal{T}}_3 \mathbf{n}_{1\bot}\cdot \hat{\mathbf{k}}\   \mathbf{n}_{2\bot} \cdot \hat{\mathbf{k}}
		 \nonumber \\
	 && +{\cal{T}}_4  \left\{ \mathbf{n}_{1}\cdot\hat{\mathbf{p}} \ \mathbf{n}_{2\bot} \cdot \hat{\mathbf{k}}+
		   \mathbf{n}_{2}\cdot \hat{\mathbf{p}} \  \mathbf{n}_{1\bot} \cdot \hat{\mathbf{k}} \right\},\label{P55fin}
\end{eqnarray}
with functions
\begin{eqnarray}
	{\cal{T}}_1 &=& s \left[ D_s \sin^2(\theta) + D_c  \cos^2(\theta) \right] \\
	{\cal{T}}_2 &=& -s D_s \sin^2(\theta)  \\
	{\cal{T}}_3 &=& s D_c \\
	{\cal{T}}_4 &=& 4Ms\sqrt{s} \  \textrm{Re} (G_MG_E^\star ) \cos(\theta) . \label{T4def}
\end{eqnarray}
Note that ${\cal{P}}^{55}$ is symmetric in $\mathbf{n}_1$ and $ \mathbf{n}_2$.

%%%%%%%%%%%%%%%%%%%%%%%%%%%%%%%%%%

  \section{Cartesian observables} \label{Sect7}
	
	Before discussing  the formulae above we introduce an ortho-normalized system
	of coordinates. The scattering plane with the vectors $\mathbf{p}$ and $\mathbf{k}$ 
	make up the $xz$-plane, with the $y$-axis  along the normal to the scattering plane,
	\begin{eqnarray}
	\mathbf{e}_z  &=&  \hat{\mathbf{p}}, \\
	\mathbf{e}_y  &=& \frac{1}{\sin(\theta) } ( \hat{\mathbf{p}}\times \hat{\mathbf{k}} ) , \\
	\mathbf{e}_x  &=& \frac{1}{\sin(\theta) } ( \hat{\mathbf{p}}\times \hat{\mathbf{k}} ) 
	 \times\hat{\mathbf{p}}.
\end{eqnarray}
Expressed in terms of these basis vectors 
\begin{equation}
	\hat{\mathbf{k}}= \sin(\theta) \mathbf{e}_x +\cos(\theta)	\mathbf{e}_z  .
\end{equation}

We first calculate the unpolarized differential cross-section distribution.
Combining eqs (\ref{dSig1}-\ref{dSig3}) and (\ref{P00fin}) we get 
\begin{eqnarray}
	\frac{\rd \sigma}{\rd \Omega}& =& \frac{\alpha^2}{4s^3} \frac{p}{k}{\cal{P}}^{00}
	 = \frac{\alpha^2}{4s^3} \frac{p}{k} ({\cal{T}}_2 + {\cal{T}}_3)\nonumber \\
	 &=& \frac{\alpha^2}{4s^2}\frac{p}{k}\left( D_c - D_s \sin^2(\theta)\right). \label{FinLLbar}
\end{eqnarray}

The components of the  vector-polarization  are obtained from eq.\ (\ref{P05fin}) 
by successively putting the polarization vector $\mathbf{n}_1$ equal to $\mathbf{e}_x$,   
$\mathbf{e}_y$, and  $\mathbf{e}_z$,
\begin{eqnarray}
	{\cal{P}}^{05}_y & =&{\cal{S}} =  -2Ms \sqrt{s} \sin(2\theta) \textrm{Im} (G_MG_E^\star ), \\
	 {\cal{P}}^{05}_x &=& {\cal{P}}^{05}_z =0.
\end{eqnarray}
The variable ${\cal{P}}^{05}_i$ is related to the more commonly used normalized-vector  polarization through
$P_i={\cal{P}}^{05}_i/{\cal{P}}^{00}$. For the anti-hyperons the polarization components are the same as
for hyperons.
 
The components of the tensor polarizations are obtained from eq.\ (\ref{P55fin}). The
diagonal components are
\begin{eqnarray}
	{\cal{P}}^{55}_{yy} & =&{\cal{T}}_2=  -s D_s \sin^2(\theta) ,\\
	{\cal{P}}^{55}_{xx} & =& {\cal{T}}_3 - {\cal{T}}_1= s \left[ D_c- D_s \right] \sin^2(\theta) ,\\
	{\cal{P}}^{55}_{zz} & =& {\cal{T}}_1 = s \left[ D_s\sin^2(\theta) + D_c \cos^2(\theta) \right],
\end{eqnarray}
the non-vanishing non-diagonal components 
\begin{eqnarray}
	{\cal{P}}^{55}_{xz} & =& \sin(\theta){\cal{T}}_4 =  2Ms \sqrt{s} \sin(2\theta) \textrm{Re} (G_MG_E^\star ) ,\nonumber\\
	& =& {\cal{P}}^{55}_{zx} ,
\end{eqnarray}
and the vanishing non-diagonal components 
\begin{equation}
	{\cal{P}}^{55}_{xy} = {\cal{P}}^{55}_{yx}= {\cal{P}}^{55}_{zy} = {\cal{P}}^{55}_{yz} =0. 
	\label{Pij-vanish}
\end{equation}
The commonly employed normalized-joint polarizations are related to the above ones 
by $A_{ij}={\cal{P}}^{55}_{ij}/{\cal{P}}^{00}$.

In the tensor polarizations the first index refers to the anti-hyperon (particle 2)
and the second to the hyperon (particle 1). However, because of the symmetry the 
remark is irrelevant.  

The polarizations and joint polarizations have been calculated before. We heve three articles 
in mind.
The first one \cite{DDR} contains several misprints. In the second one \cite{GTG} 
errors are corrected. However, both articles suggest a non-vanishing value 
for $A_{xy}$, in disagrement with our result in eq. (\ref{Pij-vanish}). For
the remaining vector and tensor polarizations  we agree with ref.\ \cite{GTG}. 
Note that our choice of coordinate system differs from that of \cite{GTG}. 
The third article \cite{BeJ} claims to obtain the same result as that of ref.\ \cite{GTG}.
However, it is easy to understand that if $A_{zy}$ vanishes, then by symmetry the same must be true for 
$A_{xy}$.

%
%
%%%%%%%%%%%%%%%%%%%%%%%%%%%%%%%%%%

%\newpage
  \section{Hyperon decay distributions}
  
  So far the analysis is valid for $e^+ e^- \rightarrow   \bar{\Lambda} \Lambda$ as well as 
  for $e^+ e^- \rightarrow   \bar{p} p$, but with different hadron form factors. Now, 
  we restrict ourselves to the first reaction.
  
	The decay distributions are obtained via the folding
	method of refs \cite{GF} and \cite{Czyz}. In this approach the hyperon-production
	distributions are multiplied by the hyperon-decay distributions and
	averged over the intermediate hyperon-spin directions. A factor of
	four must also be added since there are four spin combinations, as
	pointed out in ref.\ \cite{GF}. The labelling of momenta are explained in fig.\ 2.
	\begin{figure}[ht]
%\scalebox{0.38}{\includegraphics{Lambdafigaa.eps}\qquad \\ \includegraphics{Lambdafigbb.eps}  }
\begin{center}
\resizebox{0.38\textwidth}{!}
{ \includegraphics{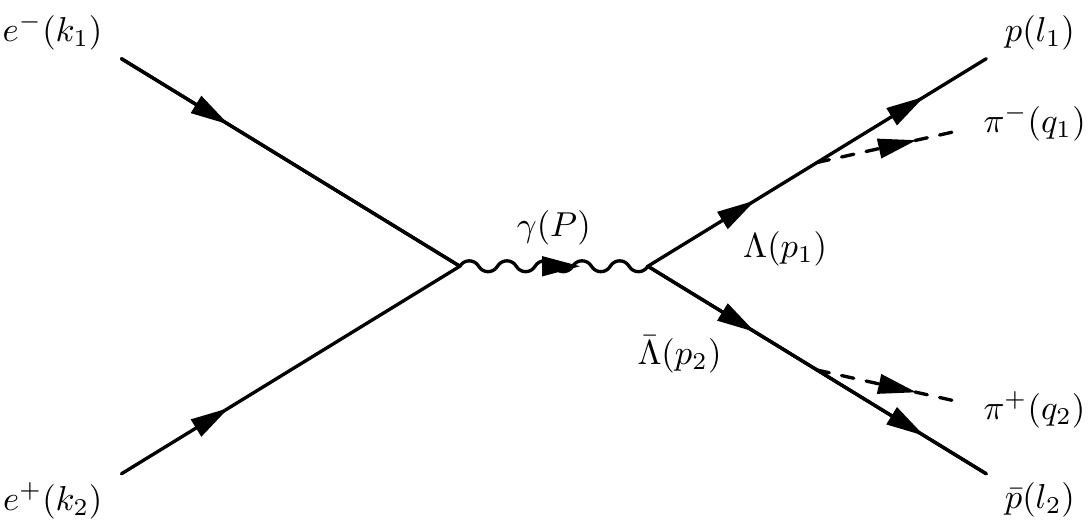}     }
\caption{Graph describing the reaction 
$e^+ e^- \rightarrow  \Lambda(\rightarrow p\pi^-)\bar{\Lambda}(\rightarrow\bar{p}\pi^+)$.}
\label{F2-fig}
\end{center}
\end{figure}

According to the folding hypothesis the distribution function is, when summed over final hadron spins,  
	\begin{align}
& \overline{|{\cal{M}}|^2} =\sum_{\pm s_1,\pm s_2}
\left\langle 	\left| {\cal{M}}(e^+ e^- \rightarrow \Lambda(s_1) \bar{\Lambda}(s_2))\right|^2 \right.
 \nonumber \\
& \qquad  \left.
\times
 \left| {\cal{M}}(\Lambda(s_1) \rightarrow  p\pi^-)\right|^2
\left| {\cal{M}}(\bar{\Lambda}(s_2) \rightarrow \bar{p}\pi^+)\right|^2  
\right\rangle_{\mathbf{n}_1\mathbf{n}_2}, \label{fold_LL}
\end{align}
with the production distribution
\begin{equation}
	\left| {\cal{M}}(e^+ e^- \rightarrow \Lambda(s_1) \bar{\Lambda}(s_2))\right|^2=L\cdot K(s_1,s_2),
\end{equation}
and the decay distribution, summed over proton spins,
	\begin{align}
& \left| {\cal{M}}(\Lambda(s_1) \rightarrow  p\pi^-)\right|^2 = R_\Lambda \big[ 1 -
	   \alpha_\Lambda l_1\cdot s_1 /l_\Lambda \big]
 \nonumber \\
& \qquad  = R_\Lambda\left[ 1 +
	   \alpha_{\Lambda} \   \mathbf{n}_1\cdot \hat{\mathbf{l}}_{1c}  \right] .
\end{align}
Here, $\mathbf{l}_{1c} $ is the proton momentum in the hyperon-rest system, local c.m.\ system,  and of length 
$| \mathbf{l}_{1c}| =l_{\Lambda}$. Expressed in 
terms of the proton momentum in the global c.m.\ system, 
\begin{align}
\mathbf{l}_{1c\bot}&=\mathbf{l}_{1\bot} , \\
	 l_{1c\parallel}&=\mathbf{l}_{1c}\cdot \hat{\mathbf{p}}= \frac{1}{p}\left[ ME_{1} -EE_{1c}\right].
\end{align}
The proton energy is $E_{1c}$ in the hyperon-rest system and $E_{1}$ in the
global c.m. \  system where the hyperon momentum is $\mathbf{p}$. The corresponding hyperon 
energies are $M$ and $E$.

The decay distribution for the anti-hyperon is similarly
\begin{align}
& \left| {\cal{M}}(\bar{\Lambda}(s_2) \rightarrow  \bar{p}\pi^+)\right|^2 = R_\Lambda \big[ 1 +
	   \alpha_\Lambda l_2\cdot s_2 /l_\Lambda \big]
 \nonumber \\
& \qquad  = R_\Lambda\left[ 1 -
	   \alpha_{\Lambda}\   \mathbf{n}_2\cdot \hat{\mathbf{l}}_{2c}   \right],
\end{align}
with
\begin{align}
\mathbf{l}_{2c\bot}&=\mathbf{l}_{2\bot} , \\
	 l_{2c\parallel}&=\mathbf{l}_{2c}\cdot \hat{\mathbf{p}}= - \frac{1}{p}\left[ ME_{2} -EE_{2c}\right].
\end{align}
Here, we have used the fact that $\alpha_{\bar{\Lambda}}=-\alpha_{\Lambda}$ and that the
anti-hyperon has momentum  $-\mathbf{p}$ in the global c.m.\ system.

The sum over the spin components in eq.\ (\ref{fold_LL}) can be replaced by a factor
of $4$ since each spin combination gives the same result when averaged 
over spin directions $\mathbf{n}_1$ and $\mathbf{n}_2$.

The averages over the spin vectors $\mathbf{n}_{1}$ and $\mathbf{n}_{2}$ in eq.\ (\ref{fold_LL})
are easily carried out since
\begin{equation}
\big{\langle } ( \mathbf{n}\cdot \mathbf{l} ) \mathbf{n}\big{\rangle }_{\mathbf{n}}   =\mathbf{l}.
 \label{N1average}
\end{equation}

The decay distribution is decomposed as,
\begin{equation}
	\overline{|{\cal{M}}|^2}= R_{\Lambda}^2 \bigg[ G^{00} +G^{05} +G^{50} + G^{55}\bigg] ,\\
	\label{MexpG}
\end{equation}
where $R_\Lambda$	is related to the decay rate of the lambda hyperon (Appendix A of \cite{GF}).
The $G$ functions are
\begin{eqnarray}
	G^{00} &=&  {\cal{P}}^{00}= {\cal{T}}_2  + {\cal{T}}_3  , \label{Gsig1} \\
	G^{05} &=&\alpha_{\Lambda}  {\cal{S}} 
	   \left[ \frac{1}{\sin(\theta)} ( \hat{\mathbf{p}}\times \hat{\mathbf{k}})\cdot \hat{\mathbf{l}}_{1c}
	      \right], \label{Gsig2}\\      
	G^{50} &=& \alpha_{\Lambda} (- {\cal{S}} )
	   \left[ \frac{1}{\sin(\theta)} ( \hat{\mathbf{p}}\times \hat{\mathbf{k}})\cdot \hat{\mathbf{l}}_{2c} 
	             \right], \label{Gsig3}\\
	G^{55} &=& \alpha_{\Lambda}^2 \bigg\{ - {\cal{T}}_1 
	  \hat{\mathbf{l}}_{1c} \cdot \hat{\mathbf{p}}\hat{\mathbf{l}}_{2c} \cdot \hat{\mathbf{p}}
	   - {\cal{T}}_2  \hat{\mathbf{l}}_{1c\bot} \cdot \hat{\mathbf{l}}_{2c\bot} \nonumber\\
		 && \quad - {\cal{T}}_3 \hat{\mathbf{l}}_{1c\bot} \cdot \hat{\mathbf{k}}\hat{\mathbf{l}}_{2c\bot} \cdot \hat{\mathbf{k}}
			\nonumber \\
			&& \quad- {\cal{T}}_4  \bigg( \hat{\mathbf{l}}_{1c} \cdot \hat{\mathbf{p}}\hat{\mathbf{l}}_{2c\bot} \cdot \hat{\mathbf{k}} 
			 + \hat{\mathbf{l}}_{2c} \cdot \hat{\mathbf{p}}\hat{\mathbf{l}}_{1c\bot} \cdot \hat{\mathbf{k}} \bigg)  \bigg\}. \label{Gsig4}
\end{eqnarray}
These equations are are expressed in terms of the proton and anti-proton momenta in the 
hyperon and anti-hyperon rest systems. However, it should be noted that $\mathbf{l}_{ic\bot}=\mathbf{l}_{i\bot} $,
for $i=1,2$.

The normalization of the cross-section distribution is as follows

\begin{eqnarray}
\frac{\rd \sigma}{\rd \Omega_{\Lambda}}  & = & \frac{1}{64 \pi^2 s}\frac{p}{k}
	  \left( \frac{4 \pi\alpha}{s}\right)^2 
		\frac{\Gamma_{\Lambda}\Gamma_{\bar{\Lambda}}}{\Gamma^2(M)}
	 \left(\sum_{a,b} G^{ab} \right)  \nonumber \\	
	&& \nonumber\\
	&& \times 
	   \left[ \frac{\rd \Omega_p}{4 \pi}\right]_{\Lambda} 
         	\left[ \frac{\rd \Omega_{\bar{p}}}{4 \pi}\right]_{\bar{\Lambda}} .\label{Sigmaster}
\end{eqnarray}
Here, $k$ and $p$	are the initial- and final-state momenta in the
reaction $e^+ e^- \rightarrow  \Lambda\bar{\Lambda}$.  The angles
$\Omega_{p}$ and $\Omega_{\bar{p}}$ are angles measured in the rest systems of $\Lambda$ 
and $\bar{\Lambda}$, and $\Omega_{\Lambda}$  in the global c.m\ system.
$\Gamma_{\Lambda}$ and $\Gamma_{\bar{\Lambda}}$ are the channel widths,

If we integrate over the lambda-decay angles $\Omega_p$  then the contributions from 
the functions $G^{05}$ and $G^{55}$ are anulled \cite{GF}, and correspondingly for 
the angles $\Omega_{\bar{p}}$. 

%
%
%%%%%%%%%%%%%%%%%%%%%%%%%%%%%%%%%%

%\newpage
  \section{Extracting form factors}
	
	The coefficients appearing in  eqs (\ref{Gsig1}-\ref{Gsig4}) are orthogonal when
	integrated over the decay distributions of the hyperons, a property which makes it 
	easy to extract the ${\cal{S}}$ and ${\cal{T}}$ functions from the cross-section distributions.
	
We first make the definitions
\begin{eqnarray}
G  & = & G^{00} +G^{05} +G^{50} + G^{55}, \\
{\rd \omega}&=&\left[ \frac{\rd \Omega_p}{4 \pi}\right]_{\Lambda} 
         	\left[ \frac{\rd \Omega_{\bar{p}}}{4 \pi}\right]_{\bar{\Lambda}}. \label{Gdefinition}
\end{eqnarray}	
With suitable weight functions we get the following extraction formulas
\begin{align}
&\int {\rd \omega}\ G = {\cal{P}}^{00} ={\cal{T}}_2 + {\cal{T}}_3,  \label{G0int}\\
&\int {\rd \omega}\ G\,  \mathbf{e}_y\cdot \hat{\mathbf{l}}_{1c} =\frac{1}{3}\  \alpha_{\Lambda}{\cal{S}},
                           \label{G1int} \\
&\int {\rd \omega}\ G\, \mathbf{e}_y\cdot \hat{\mathbf{l}}_{2c} =-\frac{1}{3}\  \alpha_{\Lambda}{\cal{S}}, 
                    \label{G2int}\\
 &\int {\rd \omega}\ G\, \hat{\mathbf{l}}_{1c} \cdot \hat{\mathbf{p}}\hat{\mathbf{l}}_{2c} \cdot \hat{\mathbf{p}}
         = - \frac{1}{9} 
                                 \  \alpha_{\Lambda}^2 {\cal{T}}_1,   \label{G3int}   \\
			 &\int {\rd \omega}\ G\,  \hat{\mathbf{l}}_{1c\bot} \cdot \hat{\mathbf{l}}_{2c\bot}  = - \frac{2}{9} 
                                 \ \alpha_{\Lambda}^2 {\cal{T}}_2,   \label{G4int}   \\													
	&\int {\rd \omega}\ G\, \hat{\mathbf{l}}_{1c\bot} \cdot \hat{\mathbf{k}}\hat{\mathbf{l}}_{2c\bot} \cdot \hat{\mathbf{k}}
	= - \frac{1}{9} \ \alpha_{\Lambda}^2 \sin^4(\theta) {\cal{T}}_3 , 
	                 \label{G5int}  \\
	&\int {\rd \omega}\ G\, \hat{\mathbf{l}}_{1c} \cdot \hat{\mathbf{p}}\hat{\mathbf{l}}_{2c\bot} \cdot \hat{\mathbf{k}} 
	= - \frac{1}{9} 
                 \ \alpha_{\Lambda}^2 \sin^2(\theta) {\cal{T}}_4.			\label{G6int}												
\end{align}	
We conclude that the $\cal{S}$ and $\cal{T}$ functions can be measured either via the vector and tensor polarizations 
 as discussed in Sect.\ \ref{Sect7}, or by  integrating  the decay-angular distributions with deftly chosen 
 weight functions.

We first integrate the measured cross section distribution $G$ over the decay distributions of both hyperons, 
eq.(\ref{G0int}), to get   
\begin{align}
	\int {\rd \omega}\ G = {\cal{T}}_2 + {\cal{T}}_3 
	=D_c\bigg[ 1-\frac{D_s}{D_c} \sin^2 (\theta)\bigg].
\end{align}
By analyzing  the angular dependence of this  distribution  
 we  determine the ratio $D_s/D_c$, which in turn yields 
the ratio of the norms of the form factors $|G_M|/|G_E|$. 
If  the absolute normalization of the cross section can be measured as well we get in addition 
the norms  of the individual form factors  $|G_M|$ and  $|G_E|$.  

After integration over both hyperon decay distributions, eq.(\ref{Sigmaster}) becomes the cross section distribution 
for the reaction $e^+ e^- \rightarrow  \Lambda\bar{\Lambda}$, eq.(\ref{FinLLbar}), but with the
extra probability factor describing the simultaneous decays $\Lambda\rightarrow p\pi^-$ and 
$\bar{\Lambda}\rightarrow \bar{p}\pi^+$,
\[ \Gamma_{\Lambda}(\Lambda\rightarrow p\pi^-) 
        \Gamma_{\bar{\Lambda}}(\bar{\Lambda}\rightarrow \bar{p}\pi^+)/{\Gamma^2(M)}.
\]
This is certainly reassuring.

The structure function called ${\cal{S}}$ contains as a factor $\textrm{Im} (G_MG_E^\star )$. 
Consequently, this factor  can be determined by integrating the measured cross section distribution 
with the projector of eq.(\ref{G1int}),
\begin{align}
&\int {\rd \omega}\ G\, \mathbf{e}_y\cdot \hat{\mathbf{l}}_{1c} =\frac{1}{3}\  \alpha_{\Lambda}{\cal{S}}
     \nonumber    \\
 & \qquad\quad= -\frac{2}{3}\, \alpha_{\Lambda}Ms \sqrt{s} \sin(2\theta) \textrm{Im} (G_MG_E^\star ),
 	\end{align}
 with  ${\cal{S}}$ from eq.(\ref{DefS}).

In order to determine  the form factors completely we also need the combination $\textrm{Re} (G_MG_E^\star)$,
which is a factor in  ${\cal{T}}_4$ of eq.(\ref{T4def}). This structure function is obtained 
with the projector  of   eq.(\ref{G6int}).
Therefore, 
\begin{align}
&\int {\rd \omega}\ G\, \hat{\mathbf{l}}_{1c} \cdot \hat{\mathbf{p}}\hat{\mathbf{l}}_{2c\bot} \cdot \hat{\mathbf{k}} 
	= - \frac{1}{9} 
                 \ \alpha_{\Lambda}^2 \sin^2(\theta) {\cal{T}}_4 \nonumber \\							
		&\qquad	= - \frac{2}{9} \ \alpha_{\Lambda}^2	Ms\sqrt{s} \   \cos(\theta)	\sin(2\theta) 	\textrm{Re} (G_MG_E^\star ).
		\label{realpdet}
	\end{align}	
 It might be appropriate to remark that if the norms of the form factors and the imaginary part as above are 
known then only the sign of the real part is free. It is determined by eq.(\ref{realpdet}).

There are four additional projections to be made. They will not give new information 
but may strengthen results already obtained. 

\appendix
%{\noname}

%{\noname}
\section{Covariant distributions}

The expressions involving the hyperon decays, eqs (\ref{Gsig1}-\ref{Gsig4}), can be made
explicitely covariant by following the methods in \cite{GF}. 

From eqs (\ref{P05}) and (\ref{P50}) we get 
\begin{align}
	{\cal{P}}^{05} (s_1,0)&= \frac{4M}{p^2}\textrm{Im}(G_MG_E^\star) \nonumber \\
	 &\qquad \times (k_1-k_2) \cdot Q \epsilon(p_1,p_2,s_1,k_1) ,\\
			{\cal{P}}^{50} (0,s_2)&= \frac{4M}{p^2}\textrm{Im}(G_MG_E^\star) \nonumber \\
	 &\qquad \times (k_2-k_1) \cdot Q \epsilon(p_1,p_2,s_2,k_2).
\end{align}
It should be observed that
\begin{equation}
	\epsilon(p_1,p_2,s_2,k_2)=-\epsilon(p_1,p_2,s_2,k_1).
\end{equation}

The folding procedure of eq.\ (\ref{fold_LL}) leads to 	
\begin{align}
G^{05}& =-  \frac{\alpha_{\Lambda}}{l_{\Lambda}}
   \bigg\langle  l_1\cdot s_1  {\cal{P}}^{05} 
                                 \bigg\rangle_{\mathbf{n}_1}, \\
	G^{50}&  = \frac{\alpha_{\Lambda}}{l_{\Lambda}}
   \bigg\langle  l_2\cdot s_2  {\cal{P}}^{50} 
                                 \bigg\rangle_{\mathbf{n}_2}.														
\end{align}
The average over the directions $\mathbf{n}$ of the  spin vector $s(p,\mathbf{n})$ 
is accomplished  by
\begin{equation}
	\left\langle s^\mu (p,\mathbf{n}) s^\nu(p,\mathbf{n}) \right\rangle_{\mathbf{n}}=
	\frac{1}{M^2} p^\mu p^\nu -g^{\mu \nu },
\end{equation}
 which in turn leads to
\begin{align}
	\left\langle (s \cdot l) s  \right\rangle_{\mathbf{n}} &=
	\lambda p- l , \\
	\lambda &= \frac{p\cdot l}{M^2}=E_0/M \nonumber \\
	    & =\frac{1}{2M^2} (M^2+m^2-\mu^2),
\end{align}
where $E_0$ is the energy of the decay proton in the hyperon rest system.

This gives the results
\begin{align}
G^{05}& =\frac{\alpha_{\Lambda}}{l_{\Lambda}}\frac{4M}{p^2}\textrm{Im}(G_MG_E^\star) \nonumber \\
	 &\qquad \times (k_1-k_2) \cdot Q \epsilon(p_1,p_2,l_1,k_1) , \\
	G^{50}&  =- \frac{\alpha_{\Lambda}}{l_{\Lambda}}\frac{4M}{p^2}\textrm{Im}(G_MG_E^\star) \nonumber \\
	 &\qquad \times (k_2-k_1) \cdot Q \epsilon(p_1,p_2,l_2,k_2).												
\end{align}
These two equations are identical to those of eqs (\ref{Gsig2}) and (\ref{Gsig3}).

We start from  eq.\ (\ref{P55}) which we expand  as 
\begin{equation}
	{\cal{P}}^{55} = A_1R_1+A_2R_2+A_3R_3+A_4R_4. \label{P55_2}
\end{equation}
The $A$ factors involve the following spin combinations
\begin{align}
	A_1=&s_1\cdot s_2, \\
	A_2=&P\cdot s_1 P\cdot s_2, \\
	A_3=&q\cdot s_1 q\cdot s_2, \\
	A_4=&P\cdot s_1 q\cdot s_2 - P\cdot s_2 q\cdot s_1,
\end{align}
with $q=k_1-k_2$, and the $R$ factors are calculted as
\begin{align}
	R_1=& s D_s \sin^2(\theta), \\
	R_2=& \frac{1}{2p^2} \bigg[ -sD_s \sin^2(\theta) -\frac{1}{2} (s+4M^2) D_c \cos^2(\theta) \nonumber \\
	 & \qquad  + 8sM^2 \cos^2(\theta)  \textrm{Re} (G_MG_E^\star ) \bigg], \\
	R_3=& D_c, \\
	R_4=& \frac{\sqrt{s}}{2p}\cos(\theta) \bigg[ D_c- 8M^2 \textrm{Re} (G_MG_E^\star ) \bigg].
\end{align}
 
We next return to the folding procedure of eq.\ (\ref{fold_LL}). We do not sum over spin
components but calculate for fixed spins, average over its directions, 
and multiply by the factor of four. The expression for the function $G^{55}$ of eq.\ (\ref{MexpG}) reads
\begin{equation}
G^{55} =- \left( \frac{\alpha_{\Lambda}}{l_{\Lambda}}\right)^2
   \bigg\langle  l_1\cdot s_1 l_2 \cdot s_2 {\cal{P}}^{55} 
                                 \bigg\rangle_{\mathbf{n}_1\mathbf{n}_2}. \label{G55aver}
\end{equation}

Let us define
\begin{equation}
	\bar{A}_i = \left\langle  l_1\cdot s_1 l_2 \cdot s_2 A_i
\right\rangle_{\mathbf{n}_1\mathbf{n}_2}.
\end{equation}
for the four possible values of $i$. 
 Then, a  straightforward calculation yields
\begin{align}
	\bar{A}_1 = & (\lambda p_1-l_1)\cdot  (\lambda p_2-l_2)          \nonumber\\
	=& - \frac{2p^2+M^2}{M^2}\ l_{1c\parallel} l_{2c\parallel} - \mathbf{l}_{1c\bot}\cdot \mathbf{l}_{2c\bot} \\
	\bar{A}_2 = & \big[ \lambda p_1\cdot P-P\cdot l_1\big]
	                \big[ \lambda p_2\cdot P-P\cdot l_2\big] \nonumber \\
		= & - \frac{sp^2}{M^2}\ l_{1c\parallel} l_{2c\parallel} , \\
		\bar{A}_3 = &\big[ \lambda p_1\cdot q-q\cdot l_1\big]
	                \big[ \lambda p_2\cdot q-q\cdot l_2\big] \nonumber \\					
									=& s\bigg[ \frac{\sqrt{s}}{2M}\cos(\theta) l_{1c\parallel}
									+ \mathbf{l}_{1c\bot}\cdot \hat{\mathbf{k}}\bigg] \nonumber \\
									&\times \bigg[  \frac{\sqrt{s}}{2M}\cos(\theta) l_{2c\parallel}
									+ \mathbf{l}_{2c\bot}\cdot \hat{\mathbf{k}}\bigg] , \\
    	\bar{A}_4 = &		\big[ \lambda p_1\cdot P-P\cdot l_1\big]	\big[ \lambda p_2\cdot q-q\cdot l_2\big] \nonumber \\
								&-\big[ \lambda p_2\cdot P-P\cdot l_2\big] \big[ \lambda p_1\cdot q-q\cdot l_1\big] \nonumber\\
								=&-\frac{ps}{M}\bigg\{
								l_{1c\parallel}\bigg[ \frac{\sqrt{s}}{2M}\cos(\theta) l_{2c\parallel}	
									+ \mathbf{l}_{2c\bot}\cdot \hat{\mathbf{k}}\bigg] \nonumber \\
							&\qquad +	l_{2c\parallel} \bigg[ \frac{\sqrt{s}}{2M}\cos(\theta) l_{1c\parallel}
									+ \mathbf{l}_{1c\bot}\cdot \hat{\mathbf{k}}\bigg] 
									  \bigg\}.
\end{align}

Inserting these results into ${\cal{P}}^{55}$ of eq.\ (\ref{P55_2}), and taking the average according to
eq.\ (\ref{G55aver}), gives the result reported in eq.\ (\ref{G6int}).

	%
%
%%%%%%%%%%%%%%%%%%%%%%%%%%%%%%%%%%
  %\newpage
\begin{acknowledgments}
	%\section{Acknowledgements}
Tord Johansson spotted the problem with the known tensor-polarization 
distributions, and suggested this investigation. I thank him, Andrzej, 
Karin, and Stefan for discussions.

\end{acknowledgments}
%%%%%%%%%%%%%%%%%%%%%%%%%%%%%%%%%%%%%%%%%%%%%%%%%%%%%%%%%%%%%%%

%%%%%%%%%%%%%%%%%%%%%%%%%%%%%%%%%%

%\newpage

\end{document}